\documentclass[a4paper]{article}

\usepackage{INTERSPEECH2021}
\usepackage{tabularx}
\usepackage{url}
\usepackage{multirow}
\usepackage{balance}
\usepackage{amsmath}

\DeclareMathOperator*{\argmin}{arg\,min}

\title{Scene-aware Far-field Automatic Speech Recognition}
\name{Zhenyu Tang, Dinesh Manocha}
\address{
  University of Maryland, College Park, MD 20742, United States}
\email{zhy@umd.edu, dmanocha@umd.edu}

\begin{document}

\maketitle
\begin{abstract}
We propose a novel method for generating scene-aware training data for far-field automatic speech recognition. We use a deep learning-based estimator to non-intrusively compute the sub-band reverberation time of an environment from its speech samples. 
We model the acoustic characteristics of a scene with its reverberation time and represent it using a multivariate Gaussian distribution. 
We use this distribution to select acoustic impulse responses from a large real-world dataset for augmenting speech data. 
The speech recognition system trained on our scene-aware data consistently outperforms the system trained using many more random acoustic impulse responses on the REVERB and the AMI far-field benchmarks. 
In practice, we obtain 2.64\% absolute improvement in word error rate compared with using training data of the same size with uniformly distributed reverberation times. 
\end{abstract}

\section{Introduction}

Automatic speech recognition (ASR) is being widely deployed in many real-world scenarios via smartphones and voice assistant devices. 
A major challenge in ASR is dealing with far-field scenarios, where the speech source is at a significant distance from the microphone~\cite{haeb2020far}. 
In far-field scenarios, the speech audio signals undergo behaviors corresponding to reflections off surfaces or diffraction around obstacles in the environment. 
These effects distort the speech signal in different manners, and thereby require much more training data to achieve high accuracy and generalization. 
However, capturing and transcribing far-field speech data with sufficient acoustic variability is a challenging task due to time and cost constraints. 
In practice, much fewer transcribed far-field training data are available than clean speech corpora. 
One strategy is that instead of recording multiple speech utterances in the same room, one can choose to record the acoustic impulse responses (AIRs) and the non-speech background noise in that room. 
These recordings can then be reused offline to artificially reverberate clean speech as if it had been recorded in that same environment~\cite{ko2017study}. 
When the AIRs and noise are properly recorded, using them to create training data for ASR produces results as good as using real recorded speech for training~\cite{szoke2019building}. 

In recent years, there has been an increasing number of recorded impulse response datasets, thereby improving the performance of far-field ASR. 
One caveat in the application of these datasets is that each dataset may only involve a limited number of different acoustic environments (e.g., hundreds of AIRs recorded in very few rooms). 
In addition, different recording devices, recording techniques, and software post-processing may have been used across datasets, which can reduce the consistency of recording among these datasets. 
When creating a training set with a combination of all the AIRs collected from different datasets, domain mismatch due to reverberation/noise level and frequency distortion can easily occur. 
In general, training ASR systems using data with as much variety as possible is considered beneficial when the goal is to make the trained model generalize to various test conditions. 
However, the inclusion of mismatched data for one domain does not necessarily help improve the performance on that specific domain. 
Further, training solely with mismatched data can significantly degrade the model performance~\cite{nguyen2019,subramanian2019investigation}. 

In many scenarios, voice assistant devices (e.g., Apple HomePod, Amazon Echo, Google Home) are operating in a given or fixed indoor or room environment whose acoustic characteristics do not change frequently. 
For this reason, knowing more acoustic characteristics about the target scenario can be helpful in selecting the appropriate training data, as the devices will be used in the same environment. 
Unfortunately, in most cases we may not know in advance the detailed information about the acoustic environment in which the ASR system will be deployed. 
Even with many popular far-field ASR benchmarks, the meta-data (e.g., room types and dimensions, mic locations, etc.) is missing or not consistently labeled~\cite{szoke2019building}, making it difficult to extract useful acoustic characteristics of the scene. 

\noindent {\bf Main Results: }We present a novel approach for generating far-field ASR training data that has similar acoustic characteristics as the target scene (i.e., scene-aware), without any apriori knowledge of the groundtruth scene characteristics or any meta-data of the scene. 
We use a learning-based method to blindly estimate the scene acoustic features in terms of sub-band reverberation time from unlabeled recorded signals. 
We fit a multivariate Gaussian distribution to the predicted feature distribution and use it to draw a desired number of AIR samples that have similar reverberation characteristics. 
We then use these samples to generate a training set which is used to train an ASR model in the same environment. 
We show the benefit of our approach by extensively comparing it to alternative data strategies on two public far-field ASR benchmarks: the REVERB challenge~\cite{kinoshita2017reverb} and the AMI corpus~\cite{carletta2006announcing}. 
Our method is able to utilize a subset that is only $5\%$ of the available real-world AIRs for training while consistently outperforming results obtained by using the full set of AIRs. 
Our method also outperforms uniformly selected subsets of the same size by up to 2.64\% word error rate. 


\section{Related Work}
Modern hybrid ASR systems mainly consist of an acoustic model (AM) which maps the audio signal to phonemes or other linguistic units and a language model (LM) which estimates the sequences of words that are most likely to be spoken~\cite{haeb2020far}. 
Both parts are highly data-driven, and many approaches have been proposed to make the hybrid system robust to context and environment changes when there is a mismatch between the test data and the training data~\cite{zhang2018deep}. 
Far-field ASR problems often result in such domain mismatches, and when no suitable far-field training data is available, domain adaptation needs to be performed for training with clean speech data. 
One way is to transform the target far-field speech to clean speech by removing reverberation and noise, also known as speech enhancement~\cite{xu2013experimental}. 
When placed as a front-end to the ASR system, it can significantly improve the ASR performance without requiring more data~\cite{weninger2015speech}, although a robust enhancement network needs to be trained beforehand. 
Another idea is to perform the reverse step by corrupting clean speech with artificial reverberation and noise, also called speech augmentation. 
Many ASR methods have utilized efficient acoustic simulators to create far-field training data from clean speech with the hope that the randomized simulation configurations may partially overlap with the target domain. 
Using simulated AIRs is an inexpensive method to provide quick improvement of the ASR system~\cite{kim2017generation,tang2020improving}. 
However, even state-of-the-art acoustic simulators still do not achieve the same level of accuracy as compared to using the real-world data~\cite{ko2017study,tang2020low}. 
A more recent trend is to use autoencoders to extract the common distribution for linguistic contents among all domains of speech data and try to train the ASR system only on the common features~\cite{tang2018study}. 

The above methods aim to eliminate/offset the influence of data from a mismatched domain. 
From a different viewpoint, it is also beneficial to explicitly build some domain awareness into the speech system. 
For example, knowing the reverberation times of different scenes can help speech dereverberation algorithms to choose the best hyperparameters~\cite{wu2016reverberation}. 
In addition, this idea also works with the more recent end-to-end ASR systems, where there are no independent AM and LM components~\cite{mirsamadi2017multi}. 
Specifically, since the speech signal embeds the environmental information, it is possible to analyze the signal and distinguish them based on their acoustic characteristics. 
Giri et al.~\cite{giri2015improving} uses a non-negative matrix factorization method to estimate the AIR from the speech signal and evaluate its reverberation time and direct-to-reverberant ratio, which are concatenated to the original feature vector of speech. 
This scheme along with multi-task learning performs the best in the simulated REVERB benchmarks. 
Another DNN approach~\cite{doulaty2016automatic} optimizes the speech perturbation level of the training set for a target domain and shows improvement on an ideal synthetic test set. 
Aside from the acoustic input, a domain-general LM can also be effectively adapted to domain specific (e.g., music, navigation, shopping) LMs with improved sub-domain ASR performance~\cite{liu2021domain}. 
Apparently, domain awareness can be built into separate stages of the ASR system. 
However, very few work have effectively investigated the selective generation process of far-field ASR training data. 


\section{Scene-aware ASR Framework}
\begin{table*}[hbt!]
\caption{Real-world AIR datasets. Some contain data other than AIRs (e.g., speech, noise). We collect 3316 AIRs by combining them. }
\label{tab:datasets}
\begin{tabularx}{\linewidth}{>{\raggedright\arraybackslash\hsize=.55\hsize\linewidth=\hsize}X>{\hsize=1.45\hsize\linewidth=\hsize}Xc}
\toprule
Name                      & Number of AIRs, channels, and recording environments                                                                     & Year \\\hline
BUT Reverb Database~\cite{szoke2019building}       & 1300+ mono channel AIRs recorded in 8 rooms.                                 & 2019 \\
MIT IR Survey~\cite{traer2016statistics}             & 271 mono channel AIRs all recorded in distinct places.                       & 2016 \\
ACE Challenge~\cite{eaton2015ace}             & \{1,2,3,5,8,32\} channel AIRs recorded in 7 rooms.                           & 2015 \\
Multichannel Impulse Response Database~\cite{hadad2014multichannel} & 234 8-channel AIRs recorded in the same room with 3 levels of reverberation and different microphone array spacings. & 2014 \\
REVERB Challenge~\cite{kinoshita2017reverb}          & 24 8-channel AIRs recorded in small, medium, and large rooms.                & 2013 \\
OpenAIR~\cite{murphy2010openair}                   & Ambisonic B format AIRs recorded in over 46 (still increasing) environments. & 2010 \\
C4DM AIR database~\cite{stewart2010database}         & 468 mono or ambisonic B format AIRs recorded in 3 large environments.        & 2010 \\
Aachen impulse response~\cite{jeub2009binaural}       & 344 binaural AIRs measured with a dummy head in 5 environments.                                   & 2009 \\
RWCP Sound Scene~\cite{nakamura1999sound} & 143 multi-channel AIRs recorded in 14 rooms.                                 & 2000 
\\\bottomrule
\end{tabularx}
\end{table*}

\subsection{Overview}
\begin{figure}[htb]
\centering 
\includegraphics[width=\linewidth]{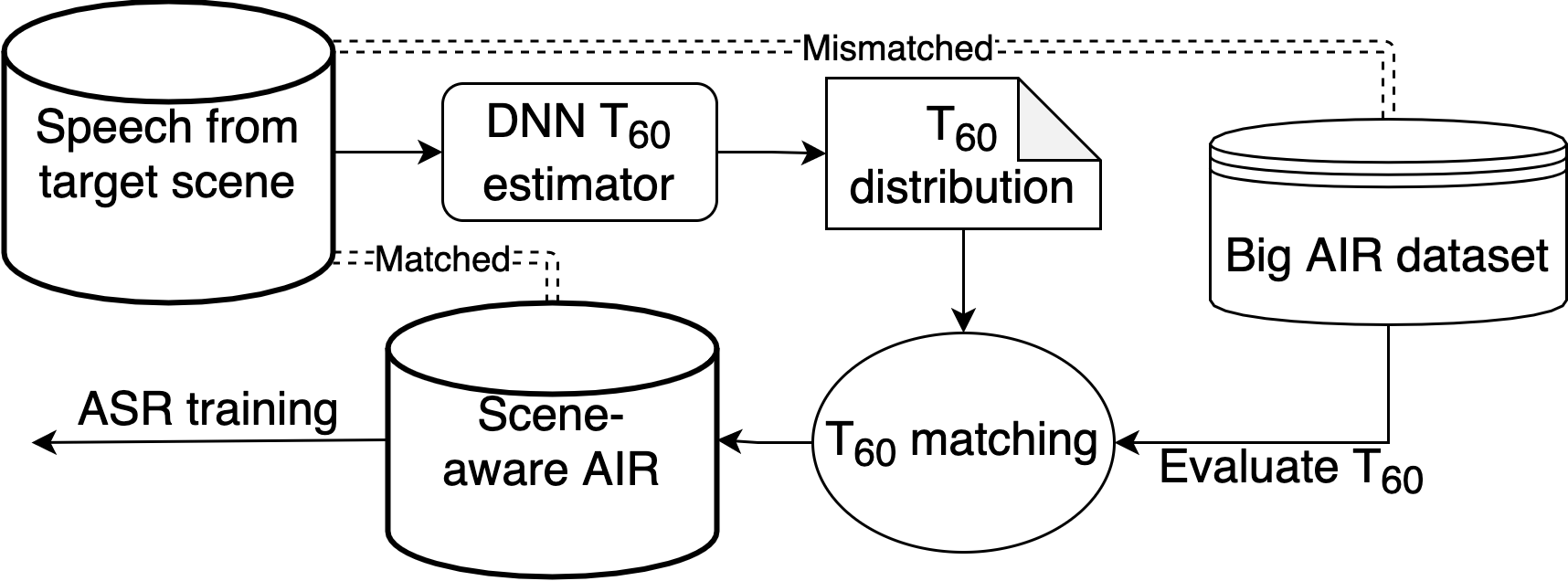}
\caption{Our scene-aware AIR selection process. When we have some speech samples from the target scene, a DNN-based $T_{60}$ estimator will help infer their distribution, modeled as a multivariate normal distribution. Then a set of scene-aware AIRs are selected from the big AIR dataset collected from the real-world according to the target distribution, which will be used for ASR training. }
\label{fig:pipeline}
\end{figure}

We consider the scenarios where clean speech is artificially reverberated using AIRs and recorded noise to create far-field training data. 
It is also more preferable to use real-world AIRs rather than simulated ones. 
When a set of AIRs and noise recordings is available, they are mixed using the formulation also described in~\cite{szoke2019building}:
\begin{equation}
\label{eq:augment}
x_{r}[t]=x[t] * h[t]+d[t],
\end{equation}
where $*$ denotes a linear convolution, $x[t]$ represents the clean speech signal, $h[t]$ is the AIR corresponding to the speech source, $d[t]$ represents ambient noise, and $x_r[t]$ represents the augmented far-field speech that is used for training. 
Obtaining a large number of $x[t]$ and $d[t]$ is relatively easy. 
Capturing $h[t]$ from the real world is a laborious work. 
Fortunately, over the past decades, multiple research groups have collected more usable AIR datasets, as listed in Table~\ref{tab:datasets}. 
Although the full set of AIRs may have more diverse reverberation characteristics than any individual dataset, we still have no guarantee that they will match the characteristics of any target scene. 
For example, if the ASR device is to be deployed in a very reverberant room, there may be many samples in the full AIR dataset that are not quite reverberant that will not help with ASR applications in these scenarios, and it may be better to train a model with matched data. 
We hypothesize that directly using the full available set of AIRs for training may not be the optimal. 
Therefore, instead of training a huge model with all the AIR augmented data and hoping it can generalize well to all scenes, we first analyze a few unlabeled speech samples from the target scene, and then select a matched set of AIRs to augment data for this scene. 
Alternatively, many ASR models that specialize in different scenarios can be pre-trained, and our scene analysis can be directly used to select a model already trained on matched data to load. 
This may be more desirable in a production setting. 
The general outline of this process is summarized in Figure~\ref{fig:pipeline}. 
In remaining parts of this section, we elaborate on how to realize such scene-awareness in a practical and non-intrusive manner. 

\subsection{Learning-based Acoustic Scene Analysis}
\label{sec:scene_analysis}
Many acoustic features can be used to describe the acoustic properties of an environment. 
The AIR fully represents the sound propagation behavior in an environment and is of crucial interest during scene analysis. 
Some standard acoustic metrics including reverberation time ($T_{60}$), direct-to-reverberant ratio (DRR), early decay time (EDT), clarity ($C_{80}$), definition ($D_{50}$), etc., are useful scalar descriptors that can be calculated from an AIR~\cite{kuttruff2016room,schissler2017acoustic}. 
In our scene-aware framework, we need to blindly estimate some of these metrics from raw speech signals because the exact AIR may not be available corresponding to the test conditions.  
$T_{60}$ is defined as the time it takes for the initial impulse energy to decay by 60dB, either for full-band or sub-band. 
It is one of the metrics most frequently referred to when comparing AIRs and the $T_{60}$ estimation topic has been studied in the ACE Challenge~\cite{eaton2015ace}. 
While there are deep learning-based methods for full-band $T_{60}$ estimation~\cite{gamper2018blind,bryan2020impulse}, we aim to predict \emph{sub-band} $T_{60}$ to better capture the frequency dependency of real-world AIRs. 
The DNN from~\cite{tang2020scene} is directly used in our approach, which consists of six 2D convolutional layers followed by a fully connected layer. 
It takes a 4-second speech spectrogram as input, and outputs 7 sub-band $T_{60}$s centered at $\{125, 250, 500, 1000, 2000, 4000, 8000\}$Hz. 
The training of this estimator is based purely on synthetic AIRs generated using methods in~\cite{bryan2020impulse}, and tested on unseen real-world AIRs from the MIT IR Survey~\cite{traer2016statistics} which has more diverse $T_{60}$s. 
Note that ideally we wish to use real-world AIR for training the analysis network instead of synthetic AIRs, but due to the unbalanced $T_{60}$ distribution of available AIRs, using them for both training and testing may cause overfitting issues. 
Also, considering that our goal is not to perfectly predict the $T_{60}$, instead our goal is to get $T_{60}$ estimates with trackable error. 
Therefore, we reserve the real-world AIRs only for testing and observe a 0.23s mean test error despite the synthetic vs. real-world domain mismatch. 

\subsection{Scene Matching and Data Generation}
\label{sec:match}
We collect publicly available AIRs, as listed in Table~\ref{tab:datasets}. 
Because they come in different audio formats, we convert all AIRs to single channel forms at a 16kHz sample rate and are able to retain 3316 such real-world AIRs in total. 
This serves as our full set of AIRs for experiments. 
Then the sub-band $T_{60}$s (a vector of length 7) of each AIR is evaluated following the method in~\cite{Karjalainen2001estimation}. 
We denote the $T_{60}$ vectors of the full AIR set as $\{\vec{s}_i\}_{i=1...K}, \vec{s}_i \in \mathbb{R}^7$, where $K=3316$. 
Now that we have the sub-band $T_{60}$ label, it is straightforward to select a subset of them to match with any desired distribution. 

\begin{figure}[htb]
\centering 
\includegraphics[width=\linewidth]{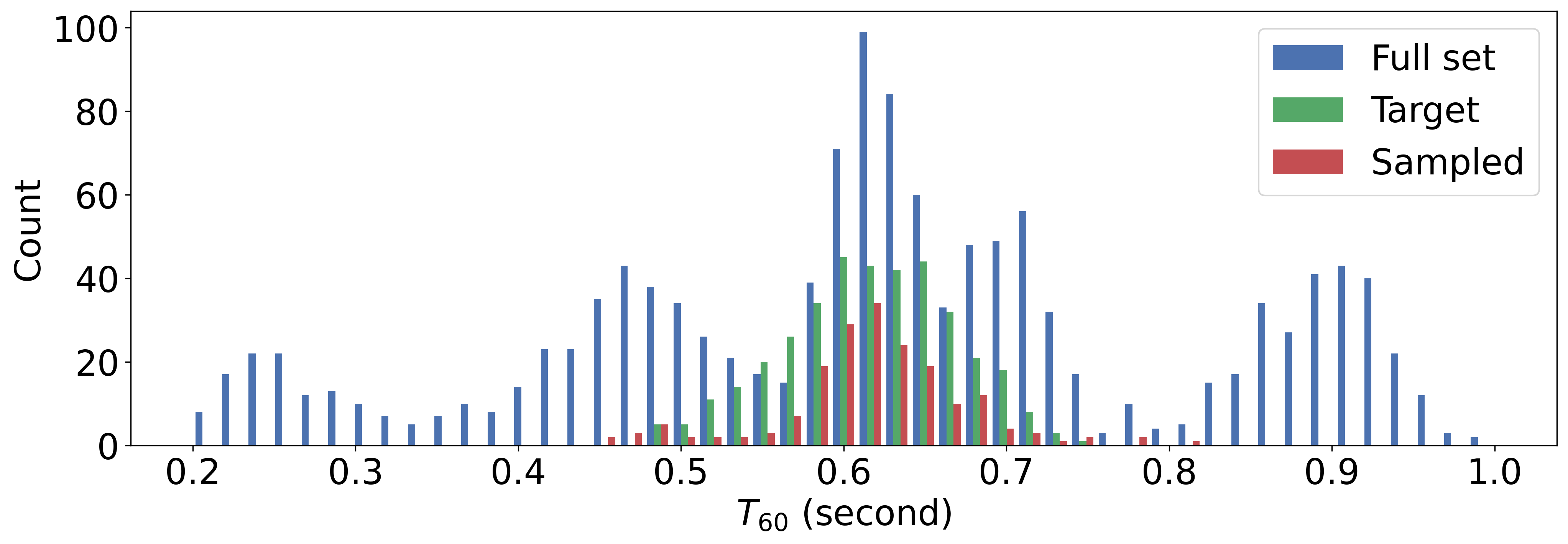}
\caption{$T_{60}$ of the full AIR set (3316 AIRs) in Table~\ref{tab:datasets} (blue, capped at 1.0s), predicted target scene $T_{60}$ distribution (green), and $T_{60}$ of the re-sampled subset of AIRs (red) according the target distribution. $T_{60}$ for the 500Hz sub-band is plotted and the results are similar for other sub-bands. The number of AIRs in the re-sampled subset is 166, but can be set to any number below the size of the full AIR set. }
\label{fig:match}
\end{figure}

Specifically, assume we have $N$ unlabeled noisy speech samples from the target scene and we want to select $M$ ($M<K$) real-world AIRs to match this scene ($M$ can be independent of $N$). 
We first use the $T_{60}$ estimator from Section~\ref{sec:scene_analysis} to create $T_{60}$ labels for them, which yields $N$ sub-band $T_{60}$ vectors $\{\vec{t}_i\}_{i=1...N}, \vec{t}_i \in \mathbb{R}^7$. 
We also observe that, while multiple measurements in the same room have some noise and the $T_{60}$ will not be identical, their $T_{60}$s resemble a Gaussian distribution. 
Therefore, we calculate the column-wise mean vector $\vec{\mu}$ and the covariance matrix $\vec{\Sigma}$ of the $N\times 7$ $T_{60}$ prediction matrix by stacking $\{\vec{t}_i\}$ vertically, and we use them to draw $M$ intermediate samples $\{\vec{\hat{t}}_i\}_{i=1...M}$ following the multivariate Gaussian distribution $\mathcal{N}(\vec{\mu}, \vec{\Sigma})$. 
Finally we can compute the pair-wise Euclidean distance between all $\{\vec{\hat{t}}_i\}$ and all $\{\vec{s}_i\}$, forming a distance matrix $D\in \mathbb{R}^{M\times K}$, where $D_{i,j}=dist(\vec{\hat{t}}_i, \vec{s}_j)$. 
Now the scene matching problem essentially becomes
\begin{equation}
\label{eq:argmin}
\argmin_{k_i\in \{1...K\}}\Sigma_{i=1}^M D_{i,k_i} \quad \text{subject to: } \forall i\neq j, k_i\neq k_j,
\end{equation}
where $\{\vec{s}_{k_i}\}_{i=1...M}$ corresponds to the $M$ selected samples that minimize the total distance. 
This is a typical assignment problem, and we use the Kuhn-Munkres algorithm~\cite{crouse2016implementing} to efficiently select $M$ samples from the full AIR set. 
An example of the matching outcome is depicted in Figure~\ref{fig:match}. 
Note that in practice, the $T_{60}$ labels may not be very accurate for the entire scene, and we want to allow some margin for errors. 
So we add the empirical mean average prediction error of the $T_{60}$ estimator to the diagonal entries of the covariance matrix $\vec{\Sigma}$ before drawing the intermediate samples. 
This makes it generate a slightly wider $T_{60}$ distribution than the original fit. 
In addition, this idea can be extended to any custom distribution sampling (e.g., uniform distribution, which we also use for experiments in Section~\ref{sec:experiments}).

\section{Experiments}
\label{sec:experiments}
In order to test the effectiveness of our scene-aware ASR approach, we conduct experiments with two far-field ASR benchmarks, and report the word error rate (WER) resulting from different AIR selection strategies. 
Specifically, we are interested in how our scene-aware AIRs compare with two alternative strategies for speech augmentation purposes: 1. use all available AIRs (i.e., the full set); 2. select a subset of AIRs with a uniform $T_{60}$ distribution. 
We use the same set of additive noise recorded in the BUT Reverb Database~\cite{szoke2019building} during augmentation for all experiments. 
Both experiments are performed with the Kaldi ASR toolbox~\footnote{\url{https://kaldi-asr.org/}} using a time delay neural network (TDNN)~\cite{peddinti2015time}. 
Each TDNN is trained on a workstation with two GeForce RTX 2080 Ti graphic cards. 
Due diligence is applied to each training process and we release our experiment code for full parameter details~\footnote{\url{https://gamma.umd.edu/pro/speech/aware}}. 

\subsection{AMI Corpus}
The AMI corpus~\cite{carletta2006announcing} consists of 100 hours of meeting recordings. 
The recordings include close-talking speech recorded by individual headset microphones (IHM) and far-field speech recorded by single distant microphones (SDM). 
While the IHM data is not of anechoic quality, it still has a very high signal-to-noise ratio compared with the SDM data, so it can be considered as clean speech. 
The original Kaldi recipe treats IHM and SDM partitions as separate tasks (i.e., both training and test sets are from the same partition), so we use a modified pipeline similar to~\cite{szoke2019building} and reverberate the IHM data using Equation~(\ref{eq:augment}) as the training set and test the trained model on SDM data. 

We perform scene matching using our method described in Section~\ref{sec:match} targeting at the SDM data. 
In total, 166 real-world AIRs ($5\%$ of the full set) have been selected for augmentation of the IHM data. 
Further, because many IHM recordings have very long durations with pauses between utterances, we also perform per-segment level speech reverberation. 
To do so, we scan each recording and, whenever a continuous 3 seconds of non-speech segment is detected, the recording is split at the beginning of the silent frames to prevent inter-segment speech overlapping after adding reverberation. 
Each segment is randomly assigned an AIR from the AIR pool (either full set, uniform set, or scene-aware set) for convolution. 
The original 687 IHM recordings are split into 17749 segments, which enable us to better utilize the AIRs for augmentation. 

\begin{table}[hbt]
\caption{WER[\%] for the AMI corpus~\cite{carletta2006announcing}. Training sets are based on reverberated IHM data except the clean IHM set, which is also not tested on the reverberated IHM (hence ``-"). 
Even though our scene-aware subset is only 5\% of the full set (with 3316 AIRs), we obtain lower WER than using the full set, as well as than using the same-sized uniform subset.}
\label{tab:ami}
\centering
\begin{tabular}{@{\extracolsep{4pt}}ccccc@{}}
\toprule
\multirow{2}{*}{AIR used} & \multicolumn{2}{c}{IHM (rvb)} & \multicolumn{2}{c}{SDM}       \\\cline{2-3}\cline{4-5}
                          & dev           & eval          & dev           & eval          \\\midrule
None (clean IHM)              & -             & -             & 54.1          & 63.5          \\
Full set                  & 39.2          & 44.0          & 44.6          & 47.7          \\
Uniform                   & 35.6          & 40.6          & 43.6          & 46.5          \\
Scene-aware               & 27.9          & 30.9 & \textbf{42.7} & \textbf{45.8} \\\bottomrule
\end{tabular}
\end{table}

Test results are shown in Table~\ref{tab:ami}. 
We also provide the results for the clean IHM training set not using any AIR, whose WER is the worst, indicating it is difficult to expect good results from severely mismatched scenes (i.e., IHM vs SDM). 
Overall, our results show that using the full set of AIRs is not optimal. 
On the SDM test data, using a uniform subset of AIRs achieves $1.2\%$ (absolute) lower WER than the full set. 
In addition, our scene-aware subset achieves $1.9\%$ (absolute) lower WER than the full set, making it the best augmentation set. 

\subsection{REVERB Challenge}
The REVERB challenge~\cite{kinoshita2017reverb} is based on the WSJCAM0 corpus~\cite{robinson1995wsjcamo}, which contains 140 speakers each speaking approximately 110 utterances. 
In this challenge, 3 rooms of different sizes are used to create artificial reverberation data (simulated rooms) by convolving their AIRs with the clean WSJCAM0 speech. 
Another large room is used to record re-transmitted WSJCAM0 speech (real room). 
Microphones are placed at two distances (near and far) from the speaker in both simulated and real rooms. 
Note that the full set contains the original AIRs from the REVERB challenge as they are sourced in Table~\ref{tab:datasets}, and we want to avoid leaking any ``groundtruth" AIRs into training, so we have excluded these AIRs from the full set during this experiment. 
From the reduced full AIR set, we select 166 AIRs to match the real room scene, and mix them with 7861 clean utterances from REVERB for training. 
We also test on simulated room 3, which is said to have very similar $T_{60}$ to the real room. 

\begin{table}[hbt]
\caption{WER[\%] for the REVERB challenge~\cite{kinoshita2017reverb}. \emph{Original} uses the original AIRs provided by the challenge, which partially match the condition in each room except the real room. \emph{Full set (-)} means excluding \emph{Original} from \emph{Full set}. Our scene-aware subset (5\% of the full set) achieves the best results in the target real room.}
\label{tab:reverb}
\setlength{\tabcolsep}{0.5pt}
\begin{tabular}{@{\extracolsep{4pt}}ccccccccc@{}}
\toprule
\multirow{3}{*}{AIR used} & \multicolumn{4}{c}{Simulated room 3} & \multicolumn{4}{c}{Real room} \\
            & \multicolumn{2}{c}{near mic}      & \multicolumn{2}{c}{far mic}       & \multicolumn{2}{c}{near mic}        & \multicolumn{2}{c}{far mic}         \\
            \cline{2-3}\cline{4-5}\cline{6-7}\cline{8-9}
                          & dev   & eval  & dev   & eval & dev   & eval  & dev   & eval  \\\midrule
Original                  & \textbf{4.23}  & \textbf{4.75}  & 6.67  & 7.40 & 18.65 & 17.90 & 20.90 & 20.51 \\                          
Full set (-)                  & 6.03  & 6.73  & 7.86  & 9.01 & 15.85 & 17.21 & 15.93 & 17.45 \\
Uniform                   & 5.24  & 6.24  & 8.08  & 9.19 & 17.59 & 18.84 & 18.25 & 19.01 \\
Scene-aware & 4.62 & 5.25 & \textbf{6.65} & \textbf{7.33} & \textbf{13.79} & \textbf{16.38} & \textbf{14.83} & \textbf{16.37} \\\bottomrule
\end{tabular}
\end{table}

The results are presented in Table~\ref{tab:reverb}. 
For the real room results, using the full set outperforms using a uniform subset. 
However, our scene-aware subset still consistently performs the best on both near and far microphone tests. 
Compared to the same-sized uniform subset, ours achieves up to a $2.64\%$ (absolute) improvement in WER. 
Even though we do not intentionally match our training set to simulated room 3, the scene-aware subset still achieves the best far microphone results, and the second best near mic results. 

\section{Conclusions and Limitations}
In this paper, we propose to use a DNN-based sub-band $T_{60}$ estimator to non-intrusively analyze speech samples from a target environment, and fit a multivariate Gaussian distribution to represent $T_{60}$ distribution effectively. 
The fitted distribution is used to guide the selection of real-world AIRs, which can generate scene-aware training data for the target environment. 
With both the REVERB challenge and the AMI corpus, our proposed scene-aware AIR always results in the ASR model with the lowest word error rate. 
Our results are promising and indicate that we need to develop scene-aware solutions for a broad class of ASR problems, especially for far-field applications. 

Our current scene-aware ASR method only uses the reverberation time of the scene to model its acoustic characteristics. 
However, there are many more acoustic parameters (mentioned in Section~\ref{sec:scene_analysis}) that may be predictable by a DNN and can be used in conjunction with $T_{60}$ to more precisely describe a scene. 
Our future goal is to extract those parameters for real-world scenes~\cite{schissler2017acoustic,tang2020scene} and use them to develop better scene-aware ASR solutions. 
It should also be noted that the selection of AIRs is still upper bounded by the size and variety of the full available set of real-world AIRs. 
If the full set is extremely imbalanced and has very few samples at some $T_{60}$ values where we want to draw a lot of samples from, no selection strategy including ours can work well. 
But we expect this issue to be resolved as more and more AIR datasets are being collected.


\bibliographystyle{IEEEtran}
\balance
\bibliography{mybib}

\end{document}